\documentclass[aps,prx,reprint,10pt,twocolumn,citeautoscript,nofootinbib,superscriptaddress]{revtex4-1}
\pdfoutput=1

\usepackage{natbib}
\usepackage[english]{babel}
\usepackage{letltxmacro}
\usepackage{latexsym}
\LetLtxMacro{\ORIGselectlanguage}{\selectlanguage}
\makeatletter
\DeclareRobustCommand{\selectlanguage}[1]{%
  \@ifundefined{alias@\string#1}
    {\ORIGselectlanguage{#1}}
    {\begingroup\edef\x{\endgroup
       \noexpand\ORIGselectlanguage{\@nameuse{alias@#1}}}\x}%
}
\newcommand{\definelanguagealias}[2]{%
  \@namedef{alias@#1}{#2}%
}
\makeatother
\definelanguagealias{en}{english}
\definelanguagealias{English}{english}
\usepackage{graphicx}
\usepackage{amsmath}
\usepackage{amsfonts}
\usepackage{amssymb}
\usepackage{bm}
\usepackage{color}
\usepackage[percent]{overpic}
\usepackage{soul} 
\usepackage{amssymb}
\usepackage{wasysym}
\usepackage{dsfont}
\usepackage{float}
\usepackage[mathscr]{euscript}
\usepackage{lipsum}
\usepackage{hyperref}
\hypersetup{
    pdfstartview={FitH},    
    colorlinks=true,       
    linkcolor=blue,          
    citecolor=blue,        
    filecolor=magenta,      
    urlcolor=blue           
}

\newcommand{\pdagger}{\phantom{\dagger}}
\newcommand{\be}{\begin{equation}}
\newcommand{\ee}{\end{equation}}
\newcommand{\bea}{\begin{eqnarray}}
\newcommand{\eea}{\end{eqnarray}}

\newcommand{\mc}{\mathcal}

\newcommand{\vect}[1]{\boldsymbol{#1}}

\usepackage{graphicx}
\usepackage[colorinlistoftodos]{todonotes}
\usepackage{verbatim}
\usepackage[normalem]{ulem}

\begin{document}

\title{Polaronic mechanism of Nagaoka ferromagnetism in Hubbard models}

\author{Rhine Samajdar}
\affiliation{Department of Physics, Princeton University, Princeton, NJ 08544, USA}
\affiliation{Princeton Center for Theoretical Science, Princeton University, Princeton, NJ 08544, USA}

\author{R. N. Bhatt}
\affiliation{Department of Physics, Princeton University, Princeton, NJ 08544, USA}
\affiliation{Department of Electrical and Computer Engineering, Princeton University, Princeton, NJ 08544, USA}

\date{\today}

\begin{abstract}
The search for elusive Nagaoka-type ferromagnetism in the Hubbard model has
recently enjoyed renewed attention with the advent of a variety of experimental platforms
enabling its realization, including moir\'e materials, quantum dots, and
ultracold atoms in optical lattices. Here, we demonstrate a universal mechanism for Nagaoka ferromagnetism (that applies to both bipartite and nonbipartite
lattices) based on the formation of ferromagnetic polarons consisting of a dopant dressed with polarized spins. Using large-scale density-matrix renormalization
group calculations, we present a comprehensive study of the
ferromagnetic polaron in an electron-doped Hubbard model, establishing various polaronic properties such as its size and energetics.  Moreover, we systematically  probe the internal structure of the magnetic state---through the use of pinning
fields and three-point spin-charge-spin correlation functions---for both the
single-polaron limit and the high-density regime of interacting polarons. Our results highlight the crucial role of mobile polarons in the birth of global ferromagnetic order from local ferromagnetism and provide a unified framework to understand the development and demise of the Nagaoka-type ferromagnetic state across dopings.
\end{abstract}

\maketitle

\section{Introduction}

The Hubbard model \cite{hubbard1964electron,gutzwiller1963effect,kanamori1963electron}, a veritable workhorse for much of our modern understanding of strongly correlated quantum matter, is believed to underlie the physics of a wide variety of complex materials \cite{lee2006doping}. In its simplest form, the model describes a system of itinerant spin-$1/2$ electrons hopping on a lattice of $N$ sites with a tunneling amplitude $t$ while interacting via a local onsite potential of strength $U$. The corresponding fermionic Hamiltonian can be written as
\begin{equation}
\label{eq:hubbard}
    H^{}_0 = - t\sum_{\langle i, j\rangle, \sigma} \left(  c^\dagger_{i\sigma} c^{\pdagger}_{j \sigma} + \mathrm{h.c.}\right) + U \sum_{i}^{} n^{\pdagger}_{i\uparrow}n^{\pdagger}_{i\downarrow},
\end{equation}
where $c^\dagger_{i,\sigma}, c^{\pdagger}_{i,\sigma}$ are the creation and annihilation operators, respectively, for an electron with spin $\sigma=\{\rvert \uparrow\rangle, \rvert \downarrow\rangle\}$ on site $i$, $n^{\pdagger}_{i,\sigma}$\,$\equiv$\,$c^\dagger_{i,\sigma} c^{\pdagger}_{i,\sigma}$ denotes the associated number operator, and the sum on $\langle i,j \rangle$ runs over all pairs of nearest-neighbor (hereafter, NN) sites. In the sixty years since its proposal, the Hubbard model and its variants have been found to host a fascinatingly diverse set of quantum phases that run the gamut from magnetic states, such as antiferromagnets and topological spin liquids, to charge density waves and superconductivity \cite{arovas2022hubbard}.

\begin{figure}[tb]
    \centering
    \includegraphics[width=0.95\linewidth]{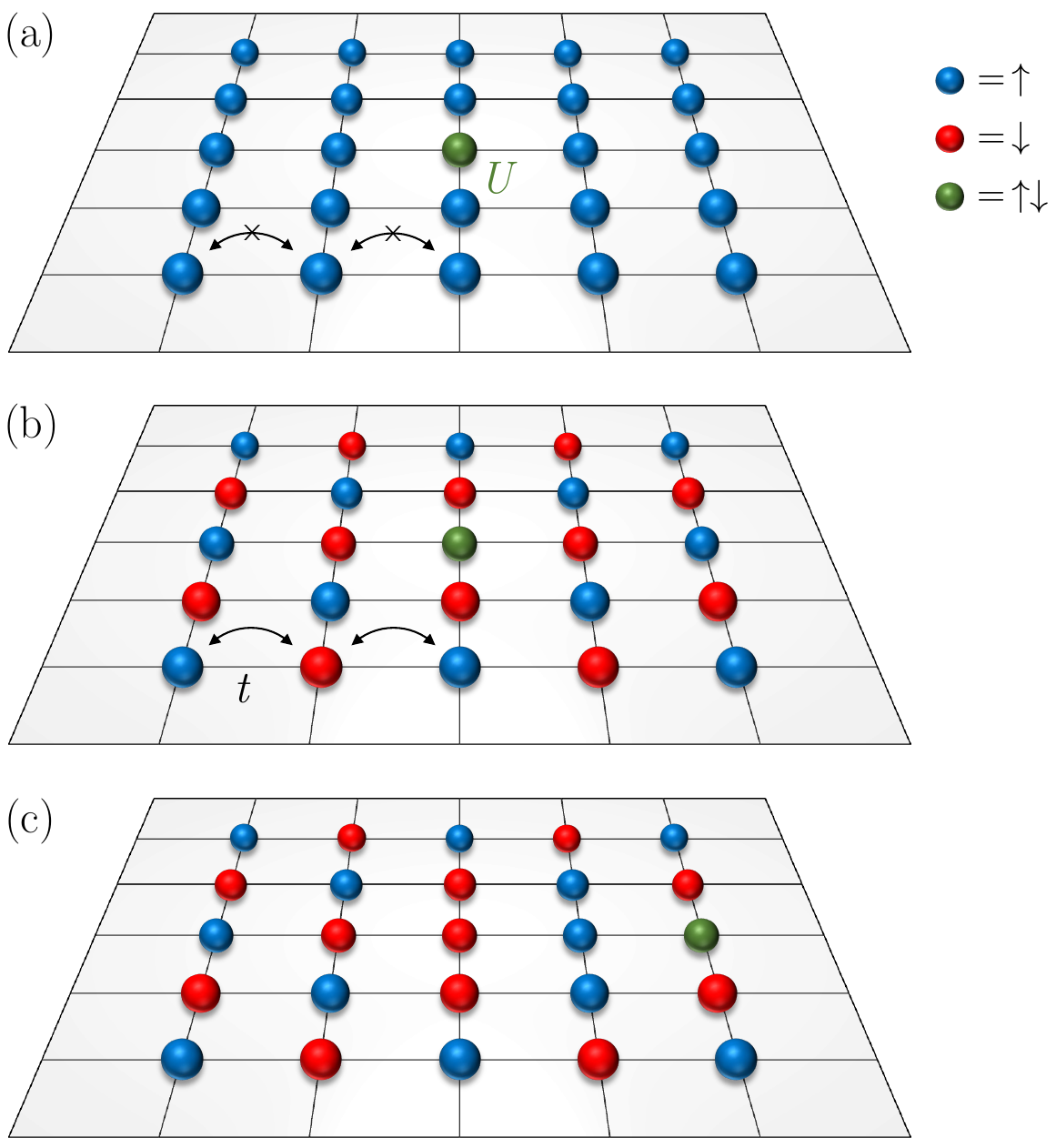}
    \caption{Schematic illustration of how ferromagnetism can be kinetically favored. (a) In a ferromagnetic state, the hopping of a down spin (red) in a background of up spins (blue) allows the doublon (green) to move freely. However, for an antiferromagnetic configuration (b), the motion of the doublon creates defects in the underlying spin texture, as sketched in (c) for a doublon moving two steps to the right. Due to the bipartite nature of the square lattice, the same argument holds irregardless of whether the relevant charge carriers are doublons or holes.}
    \label{fig:scrambling}
\end{figure}

Given the inherent complexity of the correlated electron problem, it is perhaps unsurprising that although remarkable progress has been made with numerical studies of the Hubbard model \cite{schafer2021tracking,qin2022hubbard}, to date, only a few exact analytical results are known \cite{lieb1989two,tasaki1998hubbard,li2014exact}. One such result is the rather striking Nagaoka theorem \cite{nagaoka1966ferromagnetism}, which asserts that for $U$\,$=$\,$\infty$ and nonnegative $t$, the ground state of the Hubbard model on a bipartite lattice with periodic boundary conditions (in $D$\,$\ge$\,$2$ spatial dimensions) doped with a single hole away from half filling is ferromagnetic \cite{thouless1965exchange,tasaki1989extension,tian1990simplified}, as opposed to the antiferromagnetic ground state of the half-filled system \cite{white1989numerical}. Intuitively, this follows from very general kinetic considerations, depicted in Fig.~\ref{fig:scrambling}. The hopping of dopants, either holes or doublons, necessarily scrambles an antiferromagnetic spin texture \cite{brinkman1970single,shraiman1988mobile}, leaving behind energetically unfavorable ``strings'' of displaced spins. However, such charge motion does not disrupt a ferromagnetic configuration, thus allowing carriers to be less confined, whereupon the kinetic energy gain from delocalization wins over the competing antiferromagnetic superexchange.

While mathematically rigorous, the Nagaoka theorem is of limited practical utility since any realistic system can only ever be at finite $U/t$, which introduces its own subtleties \cite{hanisch1993ferromagnetism,strack1995exact,kollar1996ferromagnetism}. Moreover, the stringent requirement of exactly one dopant is not generalizable to the thermodynamic limit; the situation with a finite density of carriers is also far from clear-cut, with arguments both for \cite{riera1989ferromagnetism,basile1990stability,hanisch1995ferromagnetism,wurth1996ferromagnetism,brunner1998quantum,becca2001nagaoka,chang2010itinerant,carleo2011itinerant,chang2011ferromagnetism} and against \cite{takahashi1982hubbard,PhysRevB.40.2719,fang1989holes,shastry1990instability,fazekas1990ground,tian1991stability,putikka1992ferromagnetism,uhrig1996exact,ivantsov2017breakdown} the existence of high-spin ground states under certain conditions.

Nonetheless, a few years ago, signatures of this elusive itinerant ferromagnetism were observed \textit{experimentally} for the first time in small (four-site) quantum dot plaquettes  \cite{dehollain2020nagaoka}. Another especially promising platform for the quantum simulation of Fermi-Hubbard models is proferred by ultracold atoms trapped in optical lattices \cite{esslinger2010fermi,von2010probing,tarruell2018quantum}, with recent experiments on these systems also demonstrating ferromagnetism \cite{xu2023frustration}, albeit in a frustrated triangular-lattice geometry. On such triangular lattices, magnetism is invariably intertwined with kinetic frustration \cite{morera2022hightemperature,schlomer2023kinetictomagnetic,samajdar2023nagaoka} as follows. As pointed out by Haerter and Shastry \cite{haerter2005kinetic}, the motion of a single hole (doublon) in a spin-polarized background leads to destructive (constructive) quantum interference between different paths on a nonbipartite lattice. To maximally lower their kinetic energy, propagating  holes therefore prefer to promote antiferromagnetic spin correlations around themselves (thereby releasing the frustration) \cite{sposetti2014classical,lisandrini2017evolution,zhang2018pairing} whereas doublons induce a local ferromagnetic environment \cite{hanisch1995ferromagnetism,kanasz2017quantum}. This phenomenon of kinetic ferromagnetism has only recently been observed in cold-atom experiments \cite{lebrat2023observation,prichard2023directly}, which demonstrated the development of ferromagnetic polarons: bound states consisting of a dopant dressed with polarized spins. A natural question to then ask, which we address in our work, is whether this mechanism of polaron formation holds even \textit{without} kinetic frustration. 

Such magnetic polarons have been extensively documented for quantum antiferromagnets in which the movement of a hole distorts the underlying N\'eel order \cite{bulaevski1968new,trugman1988interaction,schmitt1988spectral,shraiman1988mobile,sachdev1989hole,kane1989motion}. However, ferromagnetic polarons (henceforth referred to as ``Nagaoka polarons'') have been less well characterized, with nearly all theoretical studies \cite{su1988magnetic,gunn1991spin,white2001density,maska2012effective,grusdt2019microscopic} focusing on the so-called $t$-$J$ model 
[Eq.~\eqref{eq:HtJ} below], which represents an approximation to the Hubbard model in the limit of large $U/t$. Here, we present a comprehensive investigation of the Nagaoka polaron problem in a \textit{Hubbard} model, without simplification to the aforementioned $t$-$J$ limit, using large-scale density-matrix renormalization group (DMRG) calculations \cite{PhysRevLett.69.2863,PhysRevB.48.10345,RevModPhys.77.259,schollwock2011density}. In particular, we will consider an extended version of the doped Hubbard model \cite{nielsen2007nanoscale, nielsen2010search}, in which the second electron on any site of the lattice is much more weakly bound than the first, and accordingly, the hopping depends on the occupation of the site. The main advantage afforded by this model is that it greatly reduces the critical $U/t$ required for ferromagnetism on the square lattice, which is beneficial for the numerical stability of variational algorithms like DMRG. Crucially, the ferromagnetic ground states of both the extended and the regular Hubbard models belong to the same quantum phase and can be smoothly connected by  varying the microscopic parameters of the theory; hence, they share the same physics.

\begin{figure}[h]
    \centering
    \includegraphics[width=\linewidth]{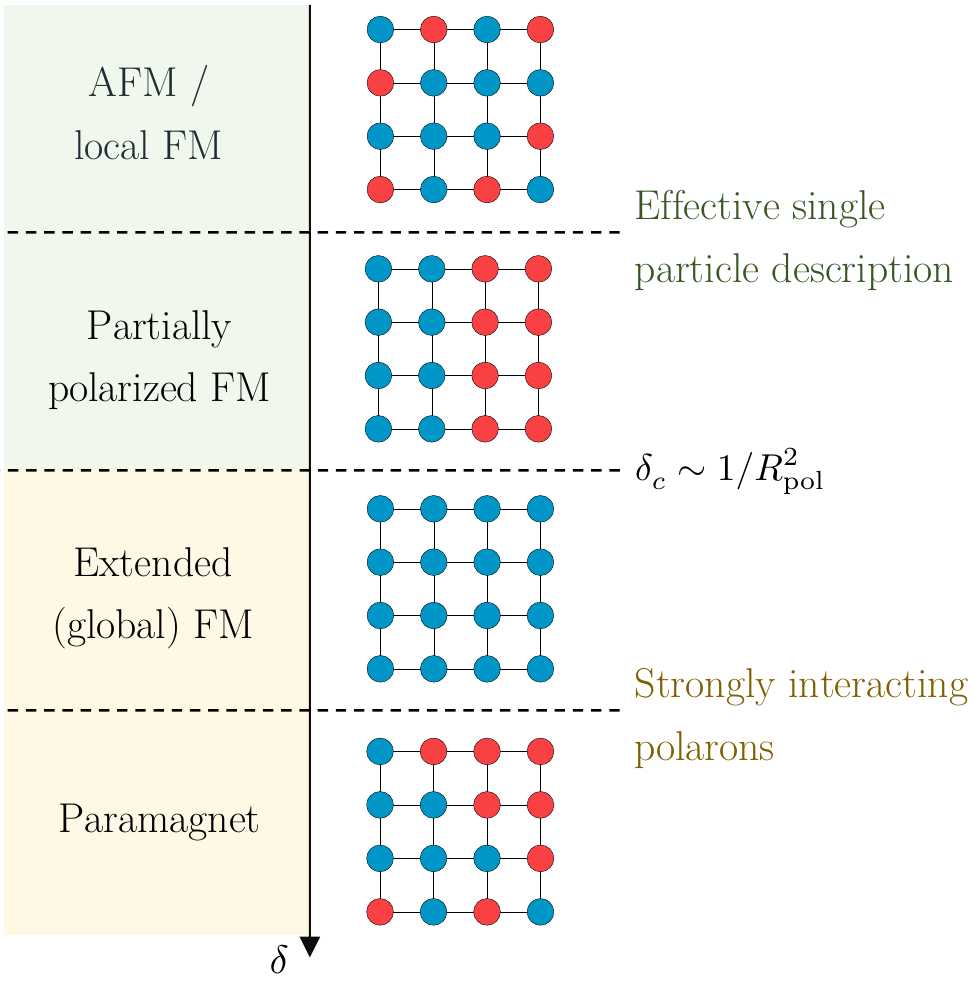}
    \caption{As a function of the electron doping concentration away from half filling, $\delta > 0$, we find four distinct magnetic regimes as illustrated pictorially here, with blue (red) sites denoting up (down) spins. For small $\delta$, the dopants form isolated polarons with local ferromagnetic (FM) ordering around each doublon core and antiferromagnetic (AFM) order further away. The radius of this polaronic cloud, $R_{\mathrm{pol}}$, grows as $U/t$ is increased. When doped with more electrons, the system crosses over to a multipolaron regime forming ferromagnetic domains (due to the kinetic energy gain from electron delocalization) but with different polarizations across domains (as favored by the superexchange). This---and the previous---regime may be viewed as a dilute gas of polarons, which is well described by an effective noninteracting  single-particle picture. As the doping is increased even further, one transitions to a regime of strongly interacting polarons at a critical $\delta_c$\,$\sim$\,$R_{\mathrm{pol}}^{-2}$; the individual mobile polarons now overlap, causing the corresponding magnetic domains to be polarized  homogeneously. The system, which can be regarded as correlated polaronic fluid in this regime, thus becomes fully ferromagnetic. Finally, at very large dopings, this global ferromagnetism is progressively destroyed due to the reduced availability of singly occupied sites, which suppresses electron hopping and fragments the extended domains. }
    \label{fig:summary}
\end{figure}

To begin, in Sec.~\ref{sec:DMRG}, we first consider square clusters with open boundary conditions and substantiate the formation of Nagaoka polarons as a route to itinerant ferromagnetism at large $U/t$. Strictly speaking, such ferromagnetism arises without all the conditions for Nagaoka's theorem being met but for the rest of this work, we adopt the nomenclature ``Nagaoka ferromagnetism'' to label this phenomenon, even though, more accurately, it is only Nagaoka-\textit{type}. We then systematically establish the properties of individual polarons---including their  energetics, size, and mobility---and discuss their extension to the higher-density regime of interacting polarons. Motivated by these observations, we turn thereafter to the study of square-lattice geometries compactified on long cylinders in Sec.~\ref{sec:cylinder}. 
With such cylindrical boundary conditions (open along the cylinder axis and periodic in the transverse direction), in addition to the fully saturated Nagaoka state, we find  striped configurations comprising ferromagnetic domains interrupted by domain walls. Irrespective of the global or local natures of the ferromagnetic order, we show the emergence of polaronic quasiparticles with various techniques, including the judicious application of pinning fields and examining specially tailored three-point spin-charge-spin correlation functions. We emphasize however that our results are obtained for finite-size systems and not necessarily for the thermodynamic limit since we cannot rule out phase separation at long length scales \cite{emery1990phase}. Our main findings are highlighted in  Sec.~\ref{sec:end} and also briefly summarized in Fig.~\ref{fig:summary}, which depicts the different magnetic ground states of the system as the doping concentration is varied.  Some additional calculations on the  one-dimensional version of this model and smaller (square) cylinders are detailed in Appendices \ref{sec:1D} and \ref{sec:CBC}, respectively.

\section{Models and methods}

The extended Hubbard model that we investigate was originally introduced to study hydrogenic donors in semiconductors [\onlinecite{nielsen2007nanoscale}, \onlinecite{nielsen2010search}; see also \onlinecite{bhatt1999monte,berciu2001effects,bhatt2002diluted}]. In an isolated hydrogen atom, the one-electron 1$s$ bound state has a binding energy of 1~Ry, but the two-electron state (H$^-$) is bound by only 0.055~Ry, i.e., it is much more weakly bound than the single-electron state. Consequently, H$^-$ is much larger in size than the neutral H atom. A Hubbard-like model for an array of hydrogenic centers thus naturally needs to be generalized to one where the hopping parameter is dependent on the occupation. This  is captured by the extended Hubbard model 
\begin{equation}
\label{eq:model}
    H = -\sum_{\langle i, j\rangle, \sigma} \left( t(n_i,n_j) c^\dagger_{i\sigma} c^{\pdagger}_{j \sigma} + \mathrm{h.c.}\right) + U \sum_i n^{\pdagger}_{i\uparrow}n^{\pdagger}_{i\downarrow},
\end{equation}
where $n_i$\,$=$\,$\sum_\sigma c^\dagger_{i,\sigma} c^{\pdagger}_{i,\sigma}$ is the total occupation of site $i$ and the correlated hopping alluded to above is
of the form
\begin{equation}
\label{eq:hopping}
    t(n^{}_i,n^{}_j) = \begin{cases}
    \Tilde{t}\,\quad &\mbox{if } n^{}_i = 1 \mbox{ and } n^{}_j = 2\\
    t \quad &\mbox{otherwise}
    \end{cases}.
\end{equation}
Note that the choice of the \textit{bare} hopping $t$ for the second case of Eq.~\eqref{eq:hopping} is essential to recover the exact asymptotic spatial dependence  \cite{herring1964asymptotic} of the effective exchange interaction $\sim$\,$e^{-2r/
a_{\textsc{b}}}$\,($\sim$\,$t^2/U$ for $t$\,$\sim$\,$e^{-r/
a_{\textsc{b}}}$), where $a_{\textsc{b}}$ is the effective Bohr radius of the hydrogenic centers. 
By construction, this model is patently electron-hole asymmetric and previously, high-spin
ground states were found to be attained at much lower $U/t$ for electron doping than hole doping  \cite{nielsen2010search}. Therefore, throughout this work, we will focus exclusively on the electron-doped case. On setting $\tilde{t} = t$, $H$ just reduces to the conventional Hubbard model $H_0$. However, a larger value of $\tilde{t}/t$ expands the regions where the ground state attains its maximum possible spin \cite{nielsen2007nanoscale} (since an enhanced hopping amplitude increases the kinetic benefit of electron delocalization). Accordingly, we will work at a fixed $\tilde{t}/t=4$ unless mentioned otherwise, but, as stressed earlier, all our conclusions about the Nagaoka polaron should apply to the  case of $\tilde{t}/t=1$  as well.

\begin{figure*}[tb]
    \centering
    \includegraphics[width=0.95\linewidth]{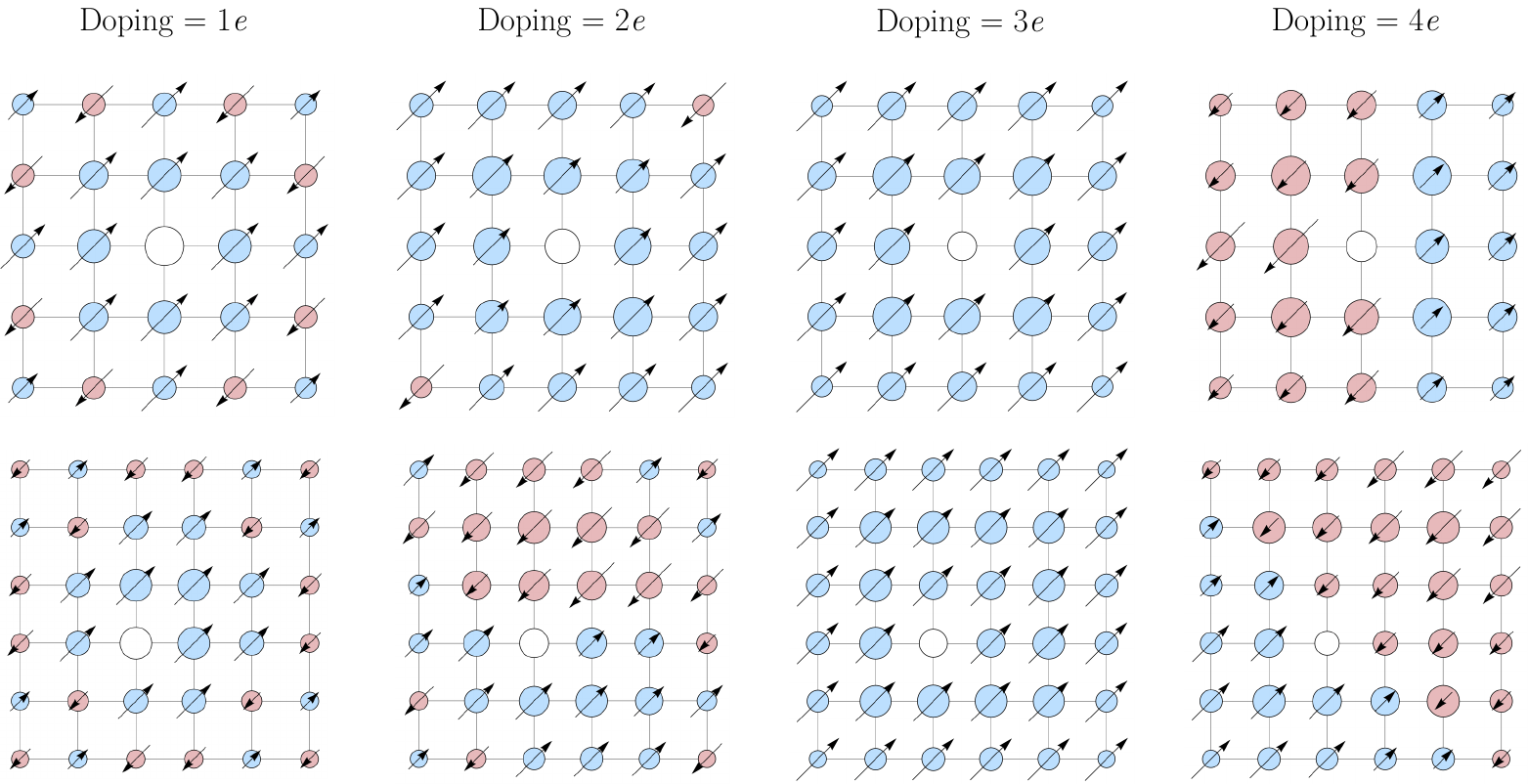}
    \caption{Ground states, with open boundary conditions, on $5\times5$ (top panel) and $6\times6$ (bottom panel)  square arrays doped with one to four electrons, for $\tilde{t}/t=4$, $U = 10\,\tilde{t}$. The diameter of each circle is proportional to the local excess electron density $\langle n_i\rangle -1$. The length of the arrows on each site indicates the magnitude of $\langle\mathbf{S}^{}_{0}\cdot\mathbf{S}^{}_{i} \rangle$, with the central (white, unmarked) site chosen as the reference spin $\textbf{S}_0$. The color of the circles as well as the orientation of the arrows conveys the sign of the spin correlations, with blue (red) denoting positive (negative) correlations.}
    \label{fig:OBC}
\end{figure*}

In the regime of large $U/t$\,$\gg$\,$1$ and at above half-filling, one can construct a low-energy theory of the extended Hubbard model \eqref{eq:model} by projecting out the unoccupied state. In this reduced Hilbert space, defined by retaining the states $\{ \rvert \uparrow \rangle, \rvert \downarrow \rangle, \rvert \uparrow \downarrow \rangle \}$ on each site, the effective Hamiltonian is  
\begin{alignat}{1}
\nonumber 
\widetilde{H} &= - \tilde{t}\hspace*{-0.1cm}\sum_{\langle i, j\rangle, \sigma}\hspace*{-0.1cm} \left(  \bar{c}^\dagger_{i\sigma} \bar{c}^{\pdagger}_{j \sigma}\hspace*{-0.1cm} + \mathrm{h.c.}\right) +\frac{4 t^2}{U} \sum_{\langle i, j\rangle} \left(\textbf{S}^{}_i\cdot\textbf{S}^{}_j - \frac{1}{4} n^{}_{i}n^{}_{j}\right),\\
S^\alpha_i &\equiv \sum_{\mu,\nu} c^\dagger_{i\mu}{\tau}^\alpha_{\mu\nu} c^{\pdagger}_{i\nu}; \quad \alpha = x,y,z,
\label{eq:HtJ1}
\end{alignat}
where $\bar{c}_{i\sigma} \equiv c_{i\sigma} n_{i\bar{\sigma}}$ is the projected electron operator ($\bar{\sigma} = \downarrow$ for $\sigma = \uparrow$ and vice versa), and $\tau^\alpha$ is a Pauli matrix in spin space.
Importantly, the spin exchange is independent of $\tilde{t}$ and equals $ 4t^2/U$ (on the square lattice), which is the same as that for the regular Hubbard model \eqref{eq:hubbard}. Defining the conventional $t$-$J$ model \cite{chao1977kinetic,chao1978canonical} as
\begin{alignat}{1}
\label{eq:HtJ}
H^{\pdagger}_{tJ} = -t\hspace*{-0.1cm} \sum_{\langle i, j\rangle, \sigma}\hspace*{-0.1cm} \left( \bar{c}^\dagger_{i\sigma} \bar{c}^{\pdagger}_{j \sigma} \hspace*{-0.05cm}+ \mathrm{h.c.}\right) \hspace*{-0.05cm}+\hspace*{-0.05cm} J \sum_{\langle i, j\rangle}\hspace*{-0.05cm} \left(\textbf{S}^{}_i\cdot\textbf{S}^{}_j - \frac{1}{4} n^{}_{i}n^{}_{j}\right),
\end{alignat}
it easy to observe that the effective Hamiltonian \eqref{eq:HtJ1} derived above is simply a rescaled version of Eq.~\eqref{eq:HtJ}, i.e., $\widetilde{H} \equiv (\tilde{t}/t) H_{tJ}$ 
with $J= 4\,t^3/(\tilde{t}\, U)$. Although we do not directly study the $t$-$J$ model in our numerical investigations, we will see that it serves as a useful descriptor of polaronic properties in certain limits.
 
We analyze the extended Hubbard model \eqref{eq:model} using DMRG, which provides an optimized  matrix product state representation of a target wavefunction. Throughout our calculations, we maintain a truncation error of $< 10^{-6}$ by adaptively increasing the bond dimension as required, up to $\chi = 6000$. Employing both open and cylindrical (unidirectionally periodic) boundary conditions, we explore the possible ground states for a broad range of $U/t$ and dopings.
In particular, we find a variety of magnetically ordered states (including fully polarized high-spin ones) at moderate to large $U/t$.

\section{Square clusters}
\label{sec:DMRG}

Most numerical studies of Nagaoka ferromagnetism in the square-lattice Hubbard model have focused on the infinite-$U$ limit, which has been investigated for small clusters \cite{dagotto1992static,chiappe1993ground,long1993hole}, with Lanczos techniques, as well as extended domains, using  dynamical mean-field theory \cite{obermeier1997ferromagnetism,park2008dynamical}, variational quantum Monte Carlo (QMC) \cite{becca2001nagaoka,ivantsov2017breakdown}, or DMRG \cite{liu2012phases,blesio2019magnetic}. However, in order to understand the magnetic interactions in the spin sector, it is important to consider the (more generalizable) Hubbard model at finite $U$. The question of ferromagnetism in this case poses a much more challenging problem, and the system sizes probed thus far have been rather limited, ranging from plaquettes of $\sim 5$--$16$ sites (amenable to exact diagonalization) \cite{nielsen2007nanoscale, nielsen2010search,buterakos2019ferromagnetism} to $\sim20$ sites in more recent works on full configuration interaction QMC \cite{yun2021benchmark,yun2023ferromagnetic}.

Here, we start by studying the full extended Hubbard model \eqref{eq:model} on $L$\,$\times$\,$L$ square arrays for $L=5,6$; our results are tabulated for clusters with open boundary conditions in Fig.~\ref{fig:OBC}. The corresponding results with cylindrical boundaries will be discussed in Appendix~\ref{sec:CBC}, Fig.~\ref{fig:CBC}.
While some ground-state properties can depend on the microscopics for these finite system sizes, let us highlight the salient features observable in Fig.~\ref{fig:OBC} that underscore a few general trends. First, on doping the system with a single electron, we can visually identify the formation of a ferromagnetic bubble residing near the center of the lattice (owing to the
boundary conditions) for both $L$\,$=$\,$5$ and $6$. While the spins are polarized within this bubble, far away from it, the spin-spin correlations turn antiferromagnetic \cite{brunner1998quantum}. By virtue of the reasoning presented in Fig.~\ref{fig:scrambling}, the doublon can move around freely only inside this bubble whereas its longer-range motion would necessarily disrupt the antiferromagnetic background. We refer to this combination of the doublon and the polarization cloud in its vicinity as a polaron.
Now, if we add an extra electron, the two clouds of polarized spins surrounding each doublon can either be of the same polarization ($L$\,$=$\,$5$) or the opposite ($L$\,$=$\,$6$). Which situation prevails is decided by the delicate interplay between the gain in kinetic energy from delocalization, which is aided by enlarging ferromagnetic domains, and the antiferromagnetic exchange energy that prefers to maximize domain walls, thereby favoring smaller  domains.  However, away from the bipolaron, the correlations still continue to be antiferromagnetic. Increasing the doping further, to three electrons, leads to the onset of long-range global ferromagnetic order that extends across the entire system (in distinction to the local ferromagnetism observed in the previous two cases) as the polaronic wavefunctions start to overlap. For both the doped $5$\,$\times$\,$5$ and $6$\,$\times$\,$6$ arrays, we find that the spins are now all fully polarized in the quantum ground state, forming a saturated Nagaoka ferromagnet. Lastly, we observe that proceeding to even higher dopant concentrations (four electrons) actually impedes ferromagnetism and the system transitions to a paramagnetic phase \cite{becca2001nagaoka}, which can be understood as follows. Recall, per Fig.~\ref{fig:scrambling}, that the very origin of ferromagnetism is due to the enhancement in the kinetic energy gained by a delocalized electron in a spin-aligned background relative to the case of the background spins being in an antiferromagnetic (or random) configuration. However, such favorable hopping processes are hindered at large  doublon concentrations because electrons cannot move between two sites which are both doubly occupied. In fact, at high densities and large $U/t$, the doublons should have correlations resembling those of free hard-core bosons \cite{long1992ground,long1993hole} and the collective charge motion is governed by a \textit{reduced} hopping probability that depends on the spin part of the wavefunction.

The ground states with cylindrical boundary conditions, shown in Fig.~\ref{fig:CBC}, are qualitatively similar, with the key difference being the development of stripes, for certain dopings, which compete with extended ferromagnetic ordering. In this section, however, we focus on arrays with open boundary conditions, deferring a detailed discussion of clusters with cylindrical boundaries to Appendix~\ref{sec:CBC}. We note that smaller ($3$\,$\times$\,$3$ and $4$\,$\times$\,$4$) clusters with fully periodic boundary conditions have been studied by Refs.~\cite{nielsen2007nanoscale, nielsen2010search}, and the results therein are in complete consistency with the polaronic mechanism that we develop below.

\subsection{Properties of the Nagaoka polaron}

Having identified the formation of magnetic polarons, we now characterize their size and energetics. To specifically study the properties of individual noninteracting polarons, we consider the case of a single electron doped into a $5$\,$\times$\,$5$ or $6$\,$\times$\,$6$ square cluster.

\begin{figure}[t]
\centering
\includegraphics[width=0.95\linewidth]{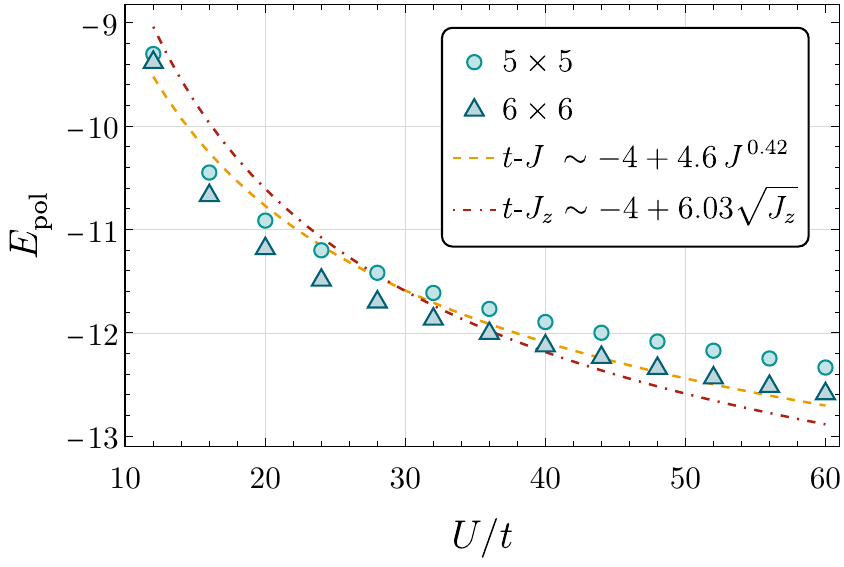}
\caption{Energy of an isolated Nagaoka polaron as a function of $U/t$ at $\tilde{t}/t=4$, as determined from Eq.~\eqref{eq:Epol} for a $5$\,$\times$\,$5$ (green circles) or $6$\,$\times$\,$6$ (blue triangles) square lattice doped with one electron above half filling. The dashed and dash-dotted lines mark the rescaled (by $\tilde{t}/t$) predictions from the $t$-$J$ [Eq.~\eqref{eq:EtJ}] and $t$-$J_z$ [Eq.~\eqref{eq:EtJz}] models, respectively.}
\label{fig:Epol}
\end{figure}

First, we compute the polaron's energy, defined as
\begin{equation}
\label{eq:Epol}
E^{}_{\mathrm{pol}} = E^{}_1 - (E^{}_0 + U), 
\end{equation}
where $E_1$ is the energy of the system doped with one excess electron and $E_0$ is that of the undoped system. $E_{\mathrm{pol}}$ therefore  represents the energy gained by ferromagnetically polarizing some subset of the spins (i.e., by the creation of the polaron) relative to the N\'eel-ordered antiferromagnetic ground state of the doublon-free undoped system. Note that in  the definition of $E_{\mathrm{pol}}$ in Eq.~\eqref{eq:Epol}, we have subtracted out a trivial  shift of the energy due to the interaction $U$ so as to isolate the magnetic contribution to the polaron's energy. For the $L$\,$=$\,$5,6$ clusters, Fig.~\ref{fig:Epol} displays that $E_{\mathrm{pol}}$ is lowered---and correspondingly, polaron formation is favored---with growing $U/t$. This behavior is in accordance with the intuition that the antiferromagnetic exchange interaction, which competes against ferromagnetism, is diminished as $U/t$ is increased.

 In the $t$-$J$ model, a straightforward analysis balancing the kinetic energy of a doublon propagating freely within a ferromagnetic droplet against the magnetic energy of the bubble (vis-à-vis the N\'eel state) shows that the energy of the polaron should scale as $\sqrt{J}$ \cite{white2001density}. Subsequent work \cite{maska2012effective} has since shown that a better numerical fit of the polaronic energy in the $t$-$J$ model is given by (in units where $t=1$)
\begin{equation}
E^{}_{tJ} = -4+4.6\, J^{\,0.42}.
\label{eq:EtJ}
\end{equation}
This curve is plotted in Fig.~\ref{fig:Epol} for comparison to our data, and the reasonable agreement of the numerically determined $E_{\mathrm{pol}}$ with this theoretical scaling further confirms our picture of polaron formation.
Furthermore, in the small-$J$ limit, the motion of a doublon is confined to its associated ferromagnetic polaron cloud, so spin-flip ($S_i^+S_j^-+S_i^-S_j^+$) processes are strongly suppressed. This motivates the consideration of an Ising version of the $t$-$J$ model \cite{dagotto1994correlated,maska2009ising}
\begin{alignat}{1}
\nonumber 
H^{\pdagger}_{tJ_z}\hspace*{-0.1cm}= -t \hspace*{-0.1cm}\sum_{\langle i, j\rangle, \sigma}\hspace*{-0.1cm}  \left(  \bar{c}^\dagger_{i\sigma} \bar{c}^{\pdagger}_{j \sigma}\hspace*{-0.1cm}  + \mathrm{h.c.}\right) + J^{}_z \sum_{\langle i, j\rangle} \left({S}^{z}_i \,{S}^{z}_j - \frac{1}{4} n^{}_{i}n^{}_{j}\right),
\end{alignat}
which drops the spin-flip part of the Heisenberg interaction in \eqref{eq:HtJ}, thereby lifting the SU($2$) spin-rotation symmetry inherent to the regular Hubbard and $t$-$J$ models.
In this case, the energy of the polaron is roughly given by \cite{white2001density}
\begin{equation}
E^{}_{tJ_z} = -4+6.03 \sqrt{J_z},
\label{eq:EtJz}
\end{equation}
which is also compared against our data in Fig.~\ref{fig:Epol}. For sufficiently low $U/t$, however, there are important corrections to the polaronic picture as described by \citet{brinkman1970single} and \citet{shraiman1988mobile}: in such a regime, the doublon can also make excursions outside the ferromagnetic bubble in the form of self-retracing walks or ``strings'', and the energy scales as $J^{2/3}$ \cite{white2001density}.

While the values of $E_{\mathrm{pol}}$ obtained for the extended Hubbard model are broadly consistent with the predictions for both $E_{tJ}$ and $E_{tJ_z}$ (after rescaling by the factor of $\tilde{t}/t$), Fig.~\ref{fig:Epol} does exhibit noticeable deviations even for large $U/t$, where the $t$-$J$ models are supposed to be good approximations. This difference between the Hubbard and $t$-$J$ behaviors can be understood by examining the higher-order magnetic interactions, which arise in a perturbative expansion of the Hubbard model. The leading correction is a biquadratic ring exchange \cite{chernyshev2004higher,reischl2004systematic} described by
\begin{alignat}{1}
\nonumber
H^{}_\square = J^{}_\square \sum_{\langle i,j,k,l \rangle} \big[ &\left(\textbf{S}^{}_i\cdot\textbf{S}^{}_j \right)\left(\textbf{S}^{}_k\cdot\textbf{S}^{}_l \right)
+ \left(\textbf{S}^{}_i\cdot\textbf{S}^{}_l \right)\left(\textbf{S}^{}_j\cdot\textbf{S}^{}_k \right) \\
-&\left(\textbf{S}^{}_i\cdot\textbf{S}^{}_k \right)\left(\textbf{S}^{}_j\cdot\textbf{S}^{}_l \right) \big],
\end{alignat}
where $i,j,k,l$ label the four spins located around a square plaquette and $J_\square$\,$\sim$\,$\mc{O}(t^4/U^3)$\,$>$\,$0$ can be as large as $20\%$ of $J$ \cite{hamerla2010derivation} depending on the bandwidth. In a ferromagnetic background, as occurring for large $U/t$, this term thus has a positive contribution, wherefore the energies of the $t$-$J$ and $t$-$J_z$ models underestimate the Hubbard $E_{\mathrm{pol}}$.

\begin{figure}[t]
\includegraphics[width=\linewidth]{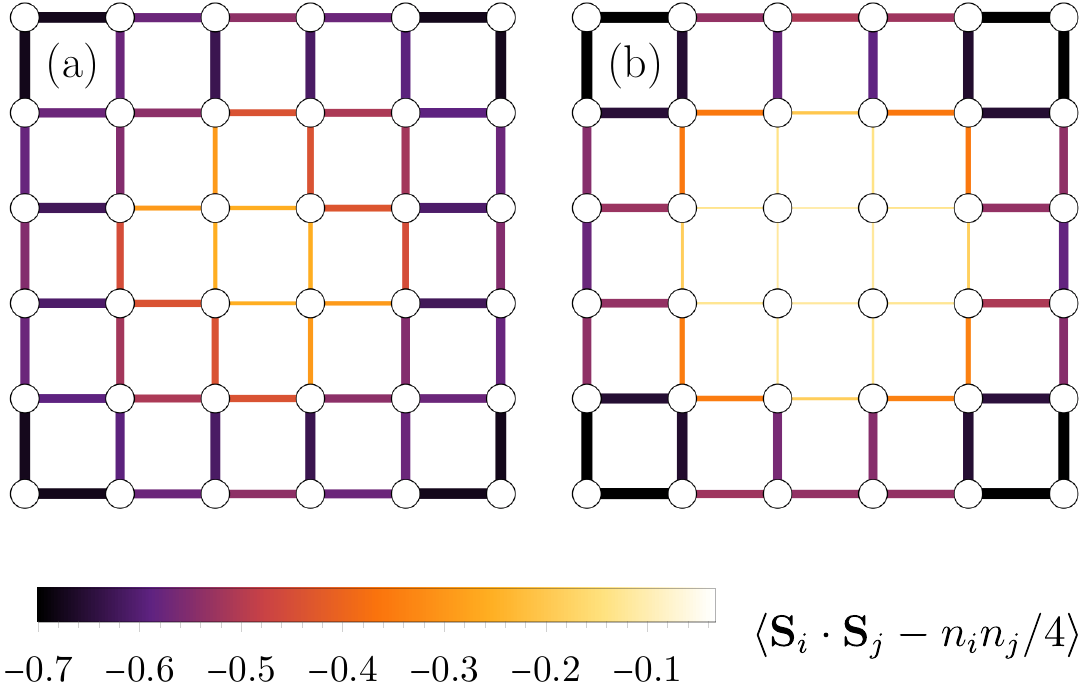}
\caption{The antiferromagnetic exchange energy $\langle\textbf{S}_i\cdot\textbf{S}_j -n_{i}n_{j}/4\rangle$ for nearest-neighboring $i$,\,$j$ on a $6$\,$\times$\,$6$ lattice doped with one excess electron at (a) $U$\,$=$\,$5\,\tilde{t}$, and (b) $U$\,$=$\,$ 10\,\tilde{t}$ ($\tilde{t}/t = 4$ in both cases). The color of each bond as well as its thickness is scaled according to the value of $\langle\textbf{S}_i\cdot\textbf{S}_j -n_{i}n_{j}/4\rangle$. The vanishing correlations at the center of the lattice delineate the extent of the ferromagnetic polaron.}
\label{fig:size}
\end{figure}

As the energy $E_{\mathrm{pol}}$ decreases with increasing $U/t$, the polaron also grows in size (as $J^{-1/4}$ for the $t$-$J$ model \cite{white2001density}), eventually expanding to fill the whole system below some threshold  $J$\,$\sim$\,$\mc{O}(1/N^2)$. This theoretically expected growth of the polaron can be observed in Fig.~\ref{fig:size}, where we plot the exchange energy $\langle\textbf{S}_i\cdot\textbf{S}_j$\,$-$\,$n_{i}n_{j}/4\rangle$ for nearest-neighboring $i,j$ on the square lattice \cite{white2001density}. From Eq.~\eqref{eq:HtJ}, one can  infer that this expectation value is a direct measure of the disturbance of an antiferromagnetic spin texture by a ferromagnetic polaron. Note that although $\langle\textbf{S}_i\cdot\textbf{S}_j$\,$-$\,$n_{i}n_{j}/4\rangle$ is indeed seen to be enhanced around a doublon in Fig.~\ref{fig:size}, it never exactly attains its maximal value of zero due to a combination of finite-size effects and the fact that we calculate this quantity for a Hubbard, rather than a $t$-$J$, model.

A similar mechanism also applies for doping with more than one electron. For a fixed $U/t$, the fraction of sites in the ferromagnetic bubble grows with the doping fraction as $R_{\mathrm{pol}}/N \propto \delta$  \cite{maska2012effective}. This consequently lowers the critical $U/t$ required for Nagaoka ferromagnetism compared to the case with a single electron dopant (as seen in Fig.~\ref{fig:OBC} above).

\subsection{Role of polaronic mobility}

Our previous calculations pertain to the limit where the density of doublons is low enough such that the system is well-described by a dilute gas of isolated polarons coupled to a spin background via the kinetic term. However, as the doping concentration is increased, interactions between these polaronic quasiparticles become more important. In this regime (see, e.g., the three- and four-electron-doped cases in Fig.~\ref{fig:OBC}), \textit{extended} ferromagnetic order can arise from the spatial overlap between the wavefunctions of different (mobile) polarons, which prompts the spins around their respective doublons to be polarized similarly. This is because if two like polarons are positioned adjacent to each other, the doublon cores of each can now  collusively delocalize over twice as large a ferromagnetic region \cite{arovas2022hubbard}.

\begin{figure}[b]
\includegraphics[width=\linewidth]{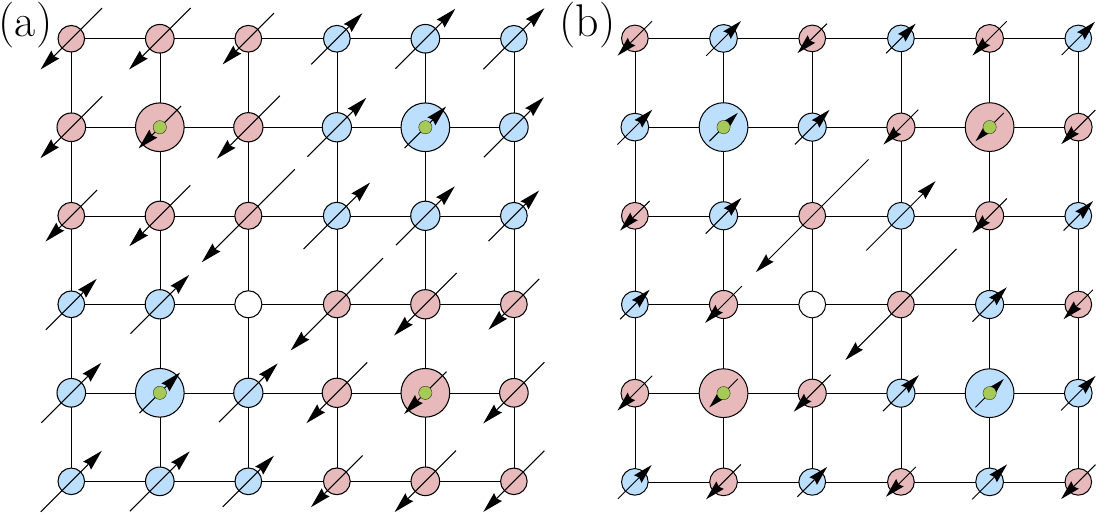}
\caption{Charge densities and spin correlations of the ground states of a $6$\,$\times$\,$6$  square lattice with open boundaries, doped with four excess electrons, for $\tilde{t}/t=4$, $U = 10\,\tilde{t}$, and pinning fields of strength (a) $V$\,$=$\,$U/4$, and (b) $V$\,$=$\,$3U/8$ applied on the four sites at the center of each $3$\,$\times$\,$3$ corner of the array (marked by green dots). }
\label{fig:pin}
\end{figure}

\begin{figure*}[tb]
\includegraphics[width=\linewidth]{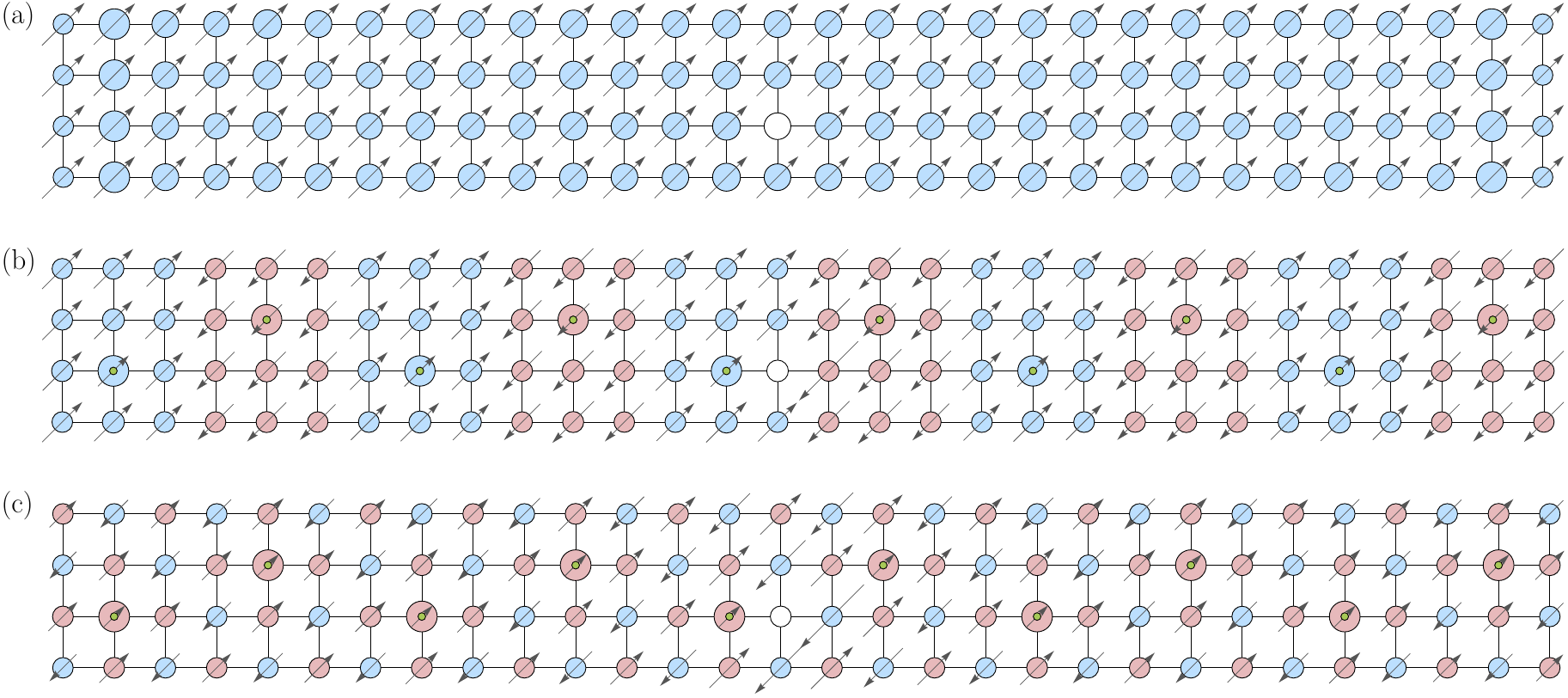}
\caption{Ground states of the extended Hubbard model \eqref{eq:model} on a $30 $\,$\times$\,$4$ cylinder at $\delta$\,$=$\,$ 1/12$ electron doping with $\tilde{t}/t=4$, $U = 10\,\tilde{t}$, and local pinning potentials (a) $V$\,$=$\,$0$, (b) $V$\,$=$\,$U/4$, and (c) $V$\,$=$\,$U/2$ applied to the ten lattice sites marked by the green dots in (b) and (c).  The effect of the pinning can be visually discerned from the growth of the charge density (the diameter of the circle) on these sites as $V$ is increased. The spin-spin correlations are plotted using the same conventions as in Fig.~\ref{fig:OBC}. }
\label{fig:cyl}
\end{figure*}

Central to this mechanism therefore is the mobility of the Nagaoka polaron. To corroborate this hypothesis, we engineer its contrapositive by explicitly pinning the doublons to certain sites of the lattice using an attractive local potential \cite{durst2002bound,koepsell2019imaging}, which disfavors their delocalization. Specifically, we consider a $6$\,$\times$\,$6$ array doped with $N_d=4$ electrons and apply a pinning potential, $-V n_s$, on four sites $s$ chosen so as to respect the rotational and reflection symmetries of the underlying square lattice.

Treating the $N_d$ doublons as spinless noninteracting fermions that fill a quadratic band, the total energy of such a multipolaron system can be easily approximated in the $t$-$J$ model, along the same  lines as the single-polaron calculation. For the optimal polaron size, this evaluates to \cite{maska2012effective}
\begin{equation}
E^{}_{tJ} (N^{}_d) = 2 N^{}_d \left(\sqrt{2\pi J}-2\right),
\label{eq:mp}
\end{equation}
in units where $t=1$.
Equation~\eqref{eq:mp} yields an initial estimate for the threshold value of the pinning potential per polaron, $V_\mathrm{th} = (\tilde{t}/t)\lvert E_{tJ} (N_d)\rvert/N_d$, that must be applied in order for the energy gain from the pinning to disrupt the ferromagnetic state.

The spin correlations of the $6$\,$\times$\,$6$ cluster (doped with four electrons) at large $U/t$ in the absence of any onsite pinning field ($V$\,$=$\,$0$) are plotted at the bottom right in Fig.~\ref{fig:OBC}. These are to be contrasted with the situation for nonzero $V$ shown  in Fig.~\ref{fig:pin}. First, upon the application of a pinning field of strength $V = U/2 < V_{\mathrm{th}}$ [Fig.~\ref{fig:pin}(a)], we see that the system forms four ferromagnetic patches, one centered around each doublon. Hence, local ferromagnetic order still persists but the domains thus formed are of smaller size than in the field-free case due to the reduced doublon mobility. On the other hand, for a potential $V$\,$=$\,$3U/8$\,$>$\,$V_{\mathrm{th}}$ [Fig.~\ref{fig:pin}(b)], this phase becomes unstable to the creation of a predominantly antiferromagnetic state but with weak ferromagnetic correlations on only the NN bonds next to the tightly pinned polarons.

\section{Polaron formation in extended systems}
\label{sec:cylinder}

Having demonstrated the origin of Nagaoka ferromagnetism via polaron formation in relatively small square arrays, we now proceed to investigate this mechanism in extended systems. To this end, we study  long
cylinders of width four and length up to $30$ sites; these dimensions are
close to the current limit of state-of-the-art ground-state DMRG numerics \cite{jiang2020ground,wietek2021stripes}. For this four-leg ladder, all our results are found to be independent of the length for cylinders that are $12$-,\,$18$-,\,$24$-, or $30$-site long, indicating that we are not limited by finite-size effects in the axial direction. We will further focus on the optimal doping fraction (for ferromagnetism) of $\delta$\,$\equiv$ $N_d/N$\,$=$\,$1/12$ suggested by the results of Fig.~\ref{fig:OBC}.

The distinctive new feature that emerges on such cylindrical geometries, as identified by Ref.~\onlinecite{samajdar2023nagaoka}, is the existence of competing magnetically ordered ground states with stripes, i.e., unidirectional charge- and spin-density modulations \cite{zaanen1989charged,machida1989magnetism,kato1990soliton}. Such an inhomogeneous striped ground state (Fig.~\ref{fig:CBC}), which breaks both rotational and translational symmetries,  arises due to the competition between the domain walls favored by the antiferromagnetic exchange and the lack thereof preferred for kinetic delocalization. However,  increasing $U/t$ weakens the spin exchange and eventually, the system undergoes a first-order quantum phase transition to a fully saturated ferromagnetic state at $U/t\sim30$  \cite{samajdar2023nagaoka}.

The polaronic nature of this ferromagnetism can be demonstrated once again by  using local pinning potentials, which we now apply in a staggered fashion along the length of the cylinder. The number of pinned sites is chosen to be the same as the number of excess electrons. When the pinning fields are absent, as mentioned above, the system exhibits long-range ferromagnetic order that spreads across the entire lattice without any domain walls [Fig.~\ref{fig:cyl}(a)]. On applying a potential of strength $V = U/4$ [Fig.~\ref{fig:cyl}(b)], this global order fractures into smaller stripes, each of a width such that it accommodates exactly one doublon on average. The natural interpretation here is that while polarons still continue to form, their extent is limited. As before, this effect of the pinning can be attributed to the reduced mobility of the doublons which, in turn, lowers the kinetic energy gain driving ferromagnetism. Upon increasing $V$ even further, to $U/2$, we observe that the polaron's radius shrinks to now encompass only the NN sites of a doublon [Fig.~\ref{fig:cyl}(c)] and the correlations in the ground state are mostly antiferromagnetic. 

\begin{figure}[t]
    \centering
\includegraphics[width=0.95\linewidth]{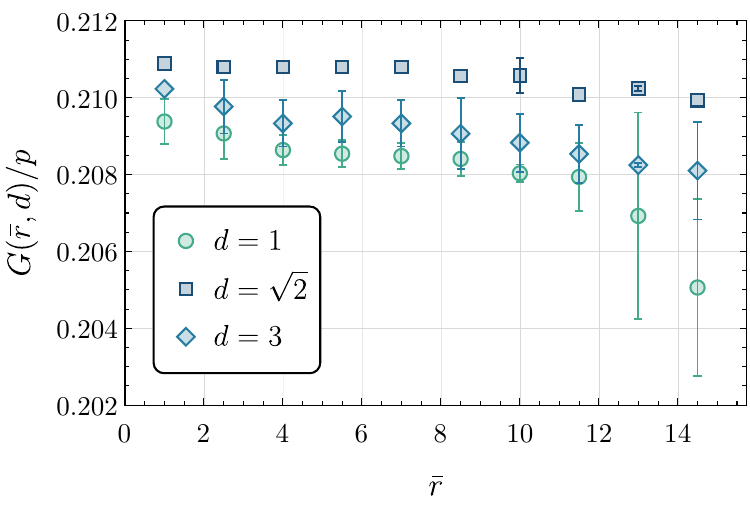}
    \caption{Decay of the (normalized) three-point correlator $G(\bar{r},d)/p$ [Eq.~\eqref{eq:3P}] in the Nagaoka ferromagnetic state of Fig.~\ref{fig:cyl}(a), as a function of the coarse-grained distance $\bar{r}$ (with $\Delta r = 0.75$) of a NN ($d=1$), 2NN ($d=\sqrt{2}$), or 3NN ($d=2$) pair of spins from the excess charge. The error bars on the $y$-axis represent the standard deviation of $G({r},d)/p$ $\forall r \in [\bar{r}-\Delta r,\bar{r}+\Delta r]$. The small dynamic range of the variation in $G$ as a function of $\bar{r}$ is indicative of the presence of long-range ferromagnetic order.}
    \label{fig:SNS}
\end{figure}

To gain further insights into the ferromagnetic polarons that we have seen develop, it is useful to probe the spin environment around the doublons at a microscopic level.
The polarization of the spins in the vicinity of a dopant electron can be quantified by a three-point function
\begin{equation}
\label{eq:3P}
    G \left(\vect{r}^{}_0; \boldsymbol{r}^{}_1,\boldsymbol{r}^{}_2\right) = \left\langle \left(n^{}_{\vect{r}_0} -1\right) \mathbf{S}^{}_{\vect{r}_1}\cdot\mathbf{S}^{}_{\vect{r}_2}\right \rangle,
\end{equation}
which measures the correlations between two spins positioned at lattice sites $\vect{r}_1$ and $\vect{r}_2$ given some excess charge density at site $\vect{r}_0$. $G \left(\vect{r}_0; \boldsymbol{r}_1,\boldsymbol{r}_2\right) $ can equivalently be expressed in terms of the displacement between the spins, $\vect{d}$\,$=$\,$\vect{r}_2$\,$-$\,$\vect{r}_1$, and the vector to the location of the doublon, $\vect{r} = (\vect{r}_1+\vect{r}_2)/2 - \vect{r}_0$, as
\begin{equation}
\label{eq:C3}
    G \left(\vect{r}^{}_0; \boldsymbol{r},\boldsymbol{d}\right) = \left\langle \left(n^{}_{\vect{r}_0} -1\right) \mathbf{S}^{}_{\vect{r}_0+\vect{r}-\vect{d}/2}\cdot\mathbf{S}^{}_{\vect{r}_0+\vect{r}+\vect{d}/2}\right \rangle.
\end{equation}
We define $G(r,d)$ as this three-point correlator spatially averaged over $\vect{r}_0$ as well as radially averaged over $\vect{r}$ and $\vect{d}$ (with $r$\,$\equiv$\,$\lvert \vect{r}\rvert$, $d$\,$\equiv$\,$\lvert \vect{d}\rvert$). Working in units where the lattice spacing $a$ is set to unity, we analyze $G(r,1)$, $G(r,\sqrt{2})$, and $G(r,2)$---which correspond to first- (NN), second- (2NN), and third-nearest-neighboring (3NN) pairs of spins, respectively---for the ferromagnetic state sketched in Fig.~\ref{fig:cyl}(a). To avoid edge effects due to the open boundaries at the ends of the cylinder, we restrict $\vect{r}_1$ and $\vect{r}_2$ to the ten central columns of the $30\times4$ lattice. Since the discrete nature of the lattice results in a set of often closely spaced distances $r$, we coarse-grain the data in (nonoverlapping) windows $[\bar{r}-\Delta r,\bar{r}+\Delta r]$ to separate out the features of the Nagaoka state, which has a long correlation length, from nonuniversal short-wavelength (lattice-scale) fluctuations. Figure~\ref{fig:SNS} plots $G(r,d)$  as a function of the distance to the doublon $r$ for three bond lengths $d$\,$=$\,$1, \sqrt{2}, 2$.  The decrease in $G(r,d)$ with increasing $r$ conveyed by Fig.~\ref{fig:SNS} indicates that spins are more likely to be aligned closer to a doublon---in consistency with our polaronic picture---while the slow nature of the decay points to the presence of long-range ferromagnetic order.

\begin{figure}[b]
    \centering
\includegraphics[width=\linewidth]{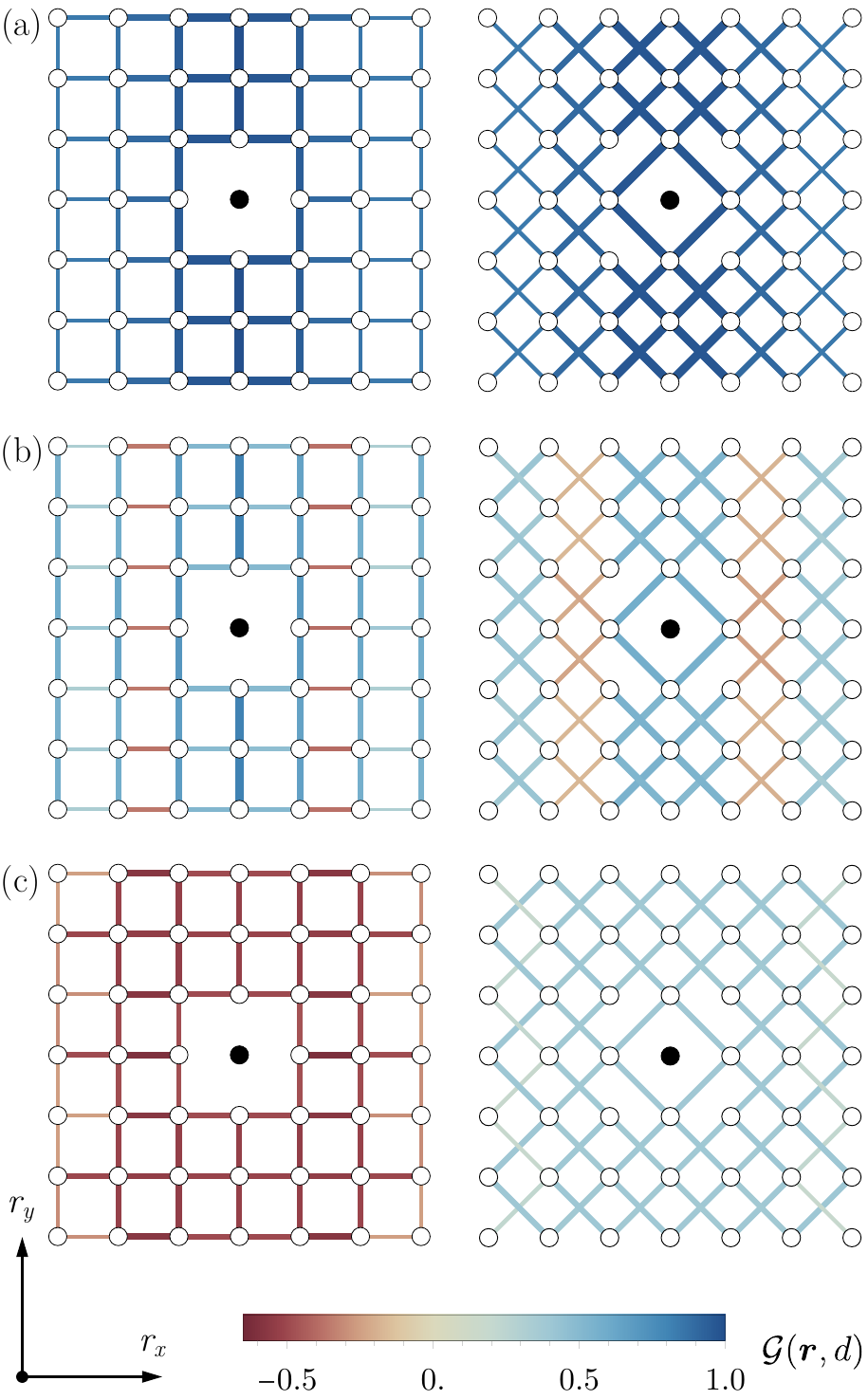}
    \caption{NN ($d$\,$=$\,$1$, left) and 2NN ($d$\,$=$\,$\sqrt{2}$, right) conditional spin correlations $\mc{G}(\vect{r},d)$, plotted as a function of $\vect{r}=(r_x,r_y)$, for the (a) ferromagnetic, (b) striped, and (c) antiferromagnetic states of Fig.~\ref{fig:cyl}. The correlations are represented by the bonds connecting two lattice sites (white dots) and are sorted according to their distance from a doublon (black circle at center).  The thickness of each bond is scaled in proportion to $\lvert \mc{G}(\vect{r},d)\rvert$. }
    \label{fig:3Pt}
\end{figure}

While $G (r,d)$ characterizes the local distortion and reorganization of magnetic correlations in the proximity of a doublon, it also includes contributions from virtual doublon-hole quantum fluctuations. Holes contribute with an opposite sign to $(n_{\vect{r}_0}-1)$ than doublons, and their effects may thus be difficult to disentangle in an averaged correlator \`a la Eq.~\eqref{eq:C3}. To circumvent this complication, we sample the ground-state DMRG wavefunction in the $\hat{z}$ basis, $\{\rvert 0 \rangle, \rvert \uparrow \rangle, \rvert \downarrow \rangle, \rvert \uparrow \downarrow \rangle \}$, and generate 100,000 snapshots; this is analogous to performing projective measurements in experiments. Using these samples, we then compute the modified three-point correlator
\begin{equation}
\label{eq:NewC3}
    \mc{G} \left(\vect{r}^{}_0; \boldsymbol{r},\boldsymbol{d}\right) = \left\langle  \mathbf{S}^{z}_{\vect{r}_0+\vect{r}-\vect{d}/2}\cdot\mathbf{S}^{z}_{\vect{r}_0+\vect{r}+\vect{d}/2}\right \rangle \bigg\rvert_{\phantom{l}^{\bullet}_\bullet \vect{r}^{}_0},
\end{equation}
which tracks the correlations between two spins separated by $\vect{d}$ conditioned on the presence of a doublon at $\vect{r}_0$  \cite{koepsell2019imaging}. Note that the quantum expectation value indicated by the angular brackets now reduces to an average over the individual sampled configurations. To differentiate between actual dopants and
naturally occurring doublon-hole fluctuations, we exclude any doubly occupied site that has a hole as its nearest neighbor. For each of the three states depicted in Fig.~\ref{fig:cyl} (ferromagnetic, striped, and antiferromagnetic), we evaluate $\mc{G} \left(\boldsymbol{r},\boldsymbol{d}\right)$---defined as $\mc{G} \left(\vect{r}_0; \boldsymbol{r},\boldsymbol{d}\right)$ averaged over all doublon positions $\vect{r}_0$---for vectors $\vect{d}$ corresponding to the NN and 2NN bonds. This spin-charge-spin correlator allows us to directly examine the internal structure of the Nagaoka polaron. For instance, in the ferromagnet [Fig.~\ref{fig:3Pt}(a)], we observe, on both NN and 2NN bonds, that the spin-spin correlations, while all positive, are strongest closest to the doublon and decay with increasing distance therefrom. Likewise, in the striped phase [Fig.~\ref{fig:3Pt}(b)], the spins immediately next to the doublon remain positively correlated. On the contrary, we find that the NN bonds situated at $r_x$\,$=$\,$\pm 1.5$ are antiferromagnetic, implying that the stripes are of width three in the $\hat{x}$ direction. The anticorrelations visible in the 2NN ($\lvert d\rvert$\,$=$\,$\sqrt{2}$) links also owe their origin to the same effect. We emphasize here that the spatial resolution of the vector $\vect{r}$ into $r_x$ and $r_y$ components proves essential for distinguishing between the ferromagnetic and striped states as the distinction between global and local ferromagnetic order can be washed out upon radially averaging $\vect{r}$. Finally, in the antiferromagnet [Fig.~\ref{fig:3Pt}(c)], we see that the NN spin-spin correlations are negative while the 2NN ones are positive. However, these (anti)correlations weaken for distances $\lvert r_x \rvert$\,$>$\,$2$, reflecting the influence of another doublon further away from the one at the origin.

\begin{figure}[t]
    \centering
\includegraphics[width=0.95\linewidth]{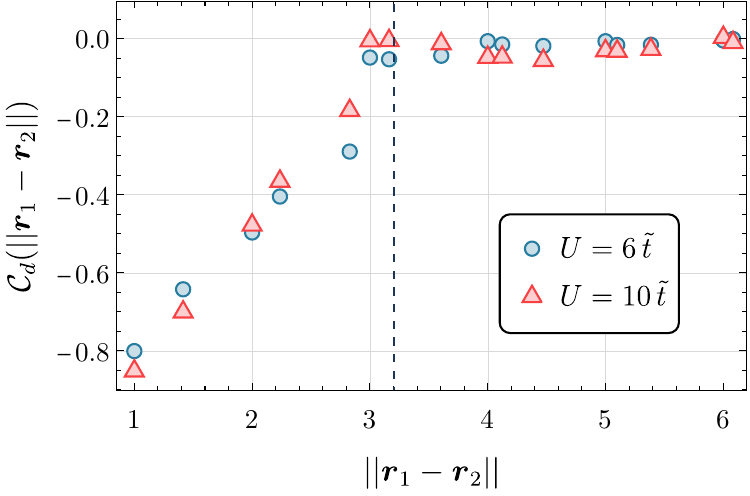}
    \caption{The radially averaged doublon-doublon correlation function of the $\delta$\,$=$\,$ 1/12$ electron-doped extended Hubbard model on a $30 $\,$\times$\,$4$ cylinder at $\tilde{t}/t=4$,  for $U$\,$=$\,$6\,\tilde{t}$ (blue circles) and $U$\,$=$\,$10\,\tilde{t}$ (red triangles), corresponding to striped and ferromagnetic ground states, respectively. The dashed line marks the average distance between doublons $\simeq 3.20$, as computed from the statistics of 100,000 projective samples of the wavefunctions. }
    \label{fig:DD}
\end{figure}

Going beyond the properties of the individual polarons established above, we can additionally probe their interplay in a multiple-dopant  system by studying the interactions between doublons. To do so, we define the doublon-doublon correlation function  \cite{koepsell2019imaging}
\begin{equation}
\label{eq:dd}
    \mc{C}^{}_d \left(\vect{r}^{}_1,\vect{r}^{}_2\right) = \frac{\left\langle n^d_{\vect{r}_1} n^d_{\vect{r}_2}\right\rangle}{\left\langle n^d_{\vect{r}_1}\right \rangle \left \langle n^d_{\vect{r}_2}\right\rangle}-1,
\end{equation}
where $n^d_{\vect{r}}=1$ if there is a doublon on site $\vect{r}$ and $0$ otherwise. Figure~\ref{fig:DD} shows that the doublons appear anticorrelated at short distances (with an exchange-correlation hole approximately three lattice spacings in size) and uncorrelated beyond this length scale, as expected for fermionic particles.

\section{Discussion and outlook}
\label{sec:end}

To summarize, in this work, we have presented extensive numerical evidence coupled with theoretical analysis to demonstrate that the formation of magnetic polarons lies at the heart of  Nagaoka ferromagnetism in the Hubbard model. Our analysis illustrates that Nagaoka ferromagnetism is fundamentally a cooperative phenomenon in that the interaction of individual polarons, each possessing only \textit{local} ferromagnetic correlations around a dopant, can engender \textit{global} ferromagnetism in a macroscopic system. This also implies that the ferromagnetism can be tuned by modifying the properties of the underlying polarons. For instance, we have seen that starting from a predominantly antiferromagnetic state, one can induce---or increase the extent of---ferromagnetic correlations by increasing $U/t$; at the microscopic level, this corresponds to enlarging the Nagaoka polaron. Conversely, given a state that \textit{is} ferromagnetic to begin with, one can destroy the long-range magnetic order by preventing the delocalization of doublons, such as via pinning potentials. This underscores the vital importance of the mobility of the polarons, which coalesce to form an extended ferromagnetic state. All these considerations taken together lead to the schematic phase diagram of Fig.~\ref{fig:summary}, which outlines the correspondence between the magnetic phases of the Hubbard model and their associated polaronic interpretations developed in our study.

Within the broader theoretical landscape, our results shed new light on the possibility and origin of itinerant ferromagnetism in the Hubbard model, a long-standing problem that has been tackled with diverse approaches over the years.
Perhaps the simplest starting point in this regard is Hartree-Fock theory, which yields ferromagnetic ground states whenever the Stoner criterion is satisfied, i.e., $D(E_F)\, U$\,$>$\,$1$, where $D(E_F)$ is the density of states at the  Fermi energy. Such a theory does predict ferromagnetism in extended regions of the Hubbard model's phase diagram, but the validity of this purely static mean-field picture expectedly breaks down in the intermediate- to strong-coupling regime where one anticipates ferromagnetism \cite{wahle1998microscopic}. A proper treatment of the Hubbard model, accounting for correlation effects, shows that the behavior of the Nagaoka ferromagnet is highly lattice-dependent \cite{hanisch1997lattice}. In particular, certain routes to ferromagnetism are often specific to nonbipartite lattices: these include the Haerter-Shastry mechanism \cite{haerter2005kinetic}, which results from frustration due to three-site loops, as well as the so-called ``low-density'' or M\"uller-Hartmann  ferromagnetism \cite{muller1995ferromagnetism} that arises due to a large and asymmetric density of states at the band edge. At first glance, this suggests that the microscopic details of the system cannot be neglected when it comes to understanding ferromagnetism, which would preclude a universal description of the physics. However, by studying the simple bipartite \textit{square} lattice here, we establish that the polaronic mechanism driving ferromagnetism is a universal and robust property of the Nagaoka state which does not rely on kinetic frustration or other lattice-specific considerations. Hence, our general conclusions regarding polaron formation should also apply to triangular lattices, which have been recently investigated in ultracold-atom experiments  \cite{xu2023frustration,lebrat2023observation,prichard2023directly} and semiconductor moir\'e superlattice systems such as WSe$_2$/WS$_2$ bilayers
 \cite{morera2022hightemperature,davydova2023itinerant,lee2023triangular}.

Looking ahead, other interesting directions in which our calculations can be extended include exploring the influence of disorder, finite temperatures, and long-ranged Coulomb interactions on ferromagnetism. Incorporation of these effects would be both useful and important for describing arrays of gate-defined semiconductor quantum dots  \cite{dehollain2020nagaoka}, which have recently emerged as another promising platform for quantum simulation of the Hubbard model \cite{hensgens2017quantum,wang2022experimental} and potentially, Nagaoka ferromagnetism.


\begin{acknowledgments}
We thank W. S. Bakr, I. Bloch, J. Dieplinger, A. Kale, L. H. Kendrick, M. Lebrat, and M. L. Prichard for useful discussions. R.S. is supported by the Princeton Quantum Initiative Fellowship. R.N.B. acknowledges support from the UK Foundation at Princeton University. This work was performed in part at the Aspen Center for Physics, which is supported by National Science Foundation grant PHY-2210452. The participation of R.S. at the Aspen Center for Physics was supported by the Simons Foundation. The calculations presented in this paper were performed using the ITensor library \cite{ITensor} on computational resources managed and supported by Princeton Research Computing, a consortium of groups including the Princeton Institute for Computational Science and Engineering (PICSciE) and the Office of Information Technology's High Performance Computing Center and Visualization Laboratory at Princeton University.

\end{acknowledgments}

\begin{appendix}

\section{One-dimensional model}
\label{sec:1D}

While the primary focus of our work has been on two spatial dimensions, the problem of ferromagnetism in the one-dimensional Hubbard model \cite{xavier2020onset} also has a long and rich history. Here, we briefly note some salient results and direct the reader to Refs.~\onlinecite{tasaki1998nagaoka,vollhardt1999metallic} for more detailed reviews. 

\begin{figure}[b]
    \centering
\includegraphics[width=\linewidth]{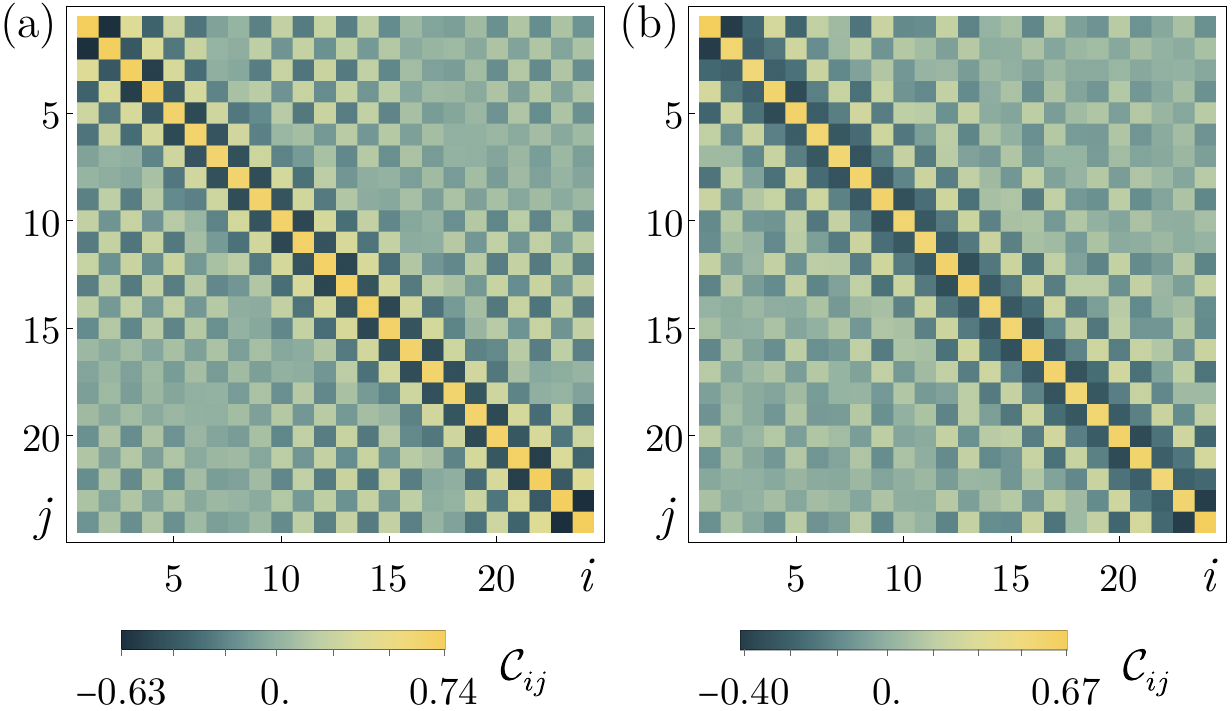}
    \caption{Two-point correlation function $\mc{C}^{}_{ij}$ of a $24$-site chain at $\tilde{t}/t$\,$=$\,$4$, $U$\,$=$\,$10\, \tilde{t}$, doped away from half filling with (a) two electrons and (b) six electrons, corresponding to dopant concentrations of $1/12$ and $1/4$, respectively. }
    \label{fig:1DCorr}
\end{figure}

For one dimension, \citet{lieb1962} rigorously proved that the  ground state of the single-band Hubbard model---with only nearest-neighbor hoppings and onsite density-density interactions---is a singlet. Therefore, obtaining ferromagnetic ground states requires circumvention of the assumptions underlying the Lieb-Mattis theorem. Broadly speaking, four different routes towards this end have been investigated.
One such way is to introduce orbital degeneracy by considering a multiband extension of the Hubbard model. Then, the local exchange interactions between electrons in different orbitals on the same site (which align unpaired electrons on each atom by Hund's rule) may lead to ferromagnetism, i.e., the hopping of holes or electrons can yield a \textit{bulk} ordering of preformed atomic moments \cite{roth1966simple, held1998microscopic,wahle1998microscopic,sakamoto2002ferromagnetism}. Another option is to add in interactions such as the nearest-neighbour Coulomb repulsion terms \cite{strack1995exact,kollar1996ferromagnetism,ueda2003ferromagnetism}, which are always present in the underlying electronic system but are abstracted away in the Hubbard model. Similarly, the inclusion of longer-range hopping  terms such as $-t_2 \sum_{i} (c^\dagger_{i,\sigma} c^{\pdagger}_{i+2,\sigma}+\mathrm{h.c.})$ has also been tied to the emergence of ferromagnetism both analytically  (in certain limits) \cite{mattis1974effect,long1994one,muller1995ferromagnetism} as well as numerically \cite{penc1996ferromagnetism,pieri1996low,daul1996dmrg, daul1998ferromagnetic,nishimoto2008phase}. This is because the proof of the Lieb-Mattis theorem relies on a definite ordering of the particles, which is no longer enforced when $ t_2 \ne 0$. Lastly, it is
possible to assemble several (identical) copies of such long-range models to obtain models with only short-range hoppings that still exhibit ferromagnetism \cite{tasaki1998nagaoka}.
This opens up the direction of inducing ``flat band'' ferromagnetism \cite{tasaki1992ferromagnetism,mielke1993,mielke1993ferromagnetism,tasaki1994stability,tasaki1995ferromagnetism,tasaki1996stability} by modifying the Hubbard model such that the lowest bands (in the single-particle spectrum) are altered to be either exactly or nearly dispersionless.

Given this backdrop, it is thus only natural to ask about the physics of the extended Hubbard model in one dimension. Rewriting the Hamiltonian \eqref{eq:model} as
\begin{alignat}{1}
\nonumber
H = &-(\tilde{t}-t)
\hspace*{-0.1cm}\sum_{\langle i, j\rangle, \sigma}  \left(c^\dagger_{i\sigma} c^{\pdagger}_{j\sigma} n^{\pdagger}_i (n^{\pdagger}_j-1) + c^\dagger_{j\sigma} c^{\pdagger}_{i\sigma} n^{\pdagger}_j (n^{\pdagger}_i-1)\right) \\
&- t\sum_{\langle i, j\rangle, \sigma} \left(  c^\dagger_{i\sigma} c^{\pdagger}_{j \sigma} + c^\dagger_{j\sigma} c^{\pdagger}_{i \sigma}\right) + U \sum_i n^{\pdagger}_{i\uparrow}n^{\pdagger}_{i\downarrow},
\end{alignat}
we observe that the hopping in the first line gets dressed by the occupation factors resulting in a four-operator term, which describes a correlated hopping process \cite{westerhout2022role} with no counterpart in the conventional Hubbard model \eqref{eq:hubbard}. Consequently, determining the ground state of this model and its spin properties  is a nontrivial task that is not immediately addressed by the Lieb-Mattis theorem.

Here, we study the one-dimensional (1D) system numerically on long chains of up to 96 sites using DMRG. The first quantity that we examine is the connected two-point correlation function,
\begin{equation}
\label{eq:C2}
    \mc{C}^{}_{ij}  = \langle  \mathbf{S}^{}_{i}\cdot\mathbf{S}^{}_{j} \rangle-\langle  \mathbf{S}^{}_{i}\rangle\cdot\langle\mathbf{S}^{}_{j} \rangle,
\end{equation}
which is plotted in Fig.~\ref{fig:1DCorr} for a chain of length $L$\,$=$\,$24$ doped away from half filling with electrons at two different doping concentrations. We observe that the dominant NN correlations are actually antiferromagnetic and upon increasing the doping fraction, the 2NN correlations also become antiferromagnetic. This  antiferromagnetic character is found to hold for a wide variety of chain lengths ($L = 12 \ell,$ $\ell=2,3,\ldots, 8$), doping concentrations ($1/12$, $1/6$, $1/4$), and model parameters ($U/\tilde{t} \in [5, 50])$, and is numerically robust in that it persists even when the system is explicitly initialized with a ferromagnetic configuration. This is in stark contrast to the behavior  in two dimensions. The difference between the two cases can be understood per the intuition outlined in Fig.~\ref{fig:scrambling}: the hopping of a dopant does not scramble an antiferromagnetic background in one dimension since the associated domain wall is a point-like (as opposed to line-like in two dimensions) object.

\begin{figure}[t]
    \centering
    \includegraphics[width=0.9\linewidth]{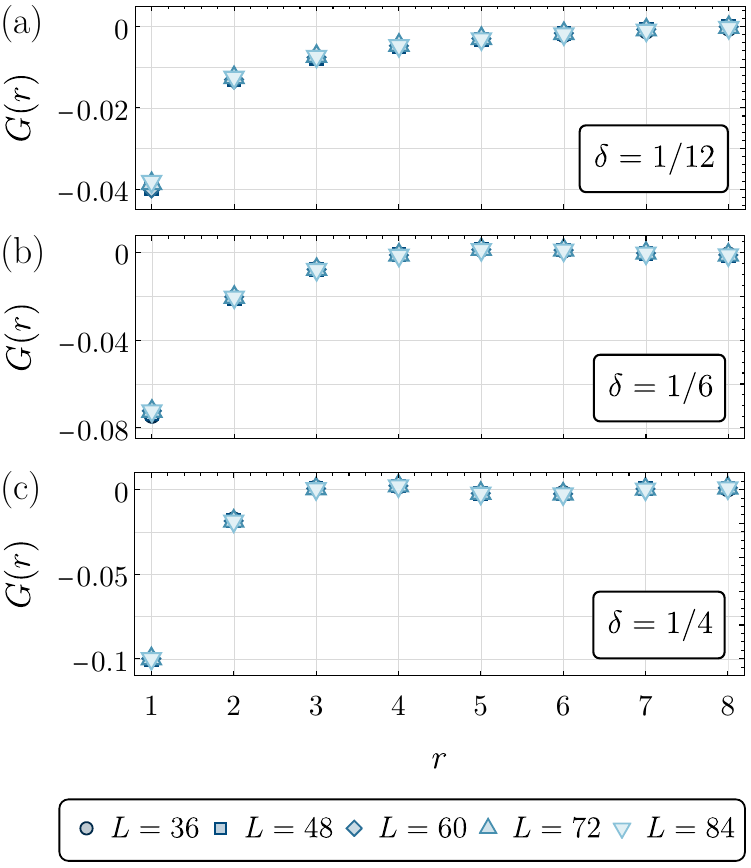}
    \caption{Absence of polaron formation in the 1D extended Hubbard model~\eqref{eq:model}. Here, we plot the three-point correlator $G^{}_i (r)$ [Eq.~\eqref{eq:C3}], averaged over the central $L/2$ lattice sites $i$, as a function of $r$ for five system sizes and electron doping concentrations (a) $1/12$, (b) $1/6$, and (c) $1/4$, at $\Tilde{t}/t = 4$, $U = 10\, \Tilde{t}$.}
    \label{fig:1Dpolaron}
\end{figure}

To microscopically probe the origin of the antiferromagnetic correlations seen in Fig.~\ref{fig:1DCorr}, it is also useful to quantify the polarization of the spins in the vicinity of a dopant electron. This is achieved by the 1D version of the three-point function in Eq.~\eqref{eq:3P},
\begin{equation}
\label{eq:C31D}
    G^{}_i (r) = \left\langle \left(n^{}_i -1\right) \mathbf{S}^{}_{i-r}\cdot\mathbf{S}^{}_{i+r} \right\rangle,
\end{equation}
which should show the development of a ferromagnetic polaron, if any. In Fig.~\ref{fig:1Dpolaron}, we plot the correlation function $G^{}_i (r)$ averaged over $i$, denoted as $G (r)$, for several different lattice sizes and doping concentrations in the regime of large $U/t$ where one might expect ferromagnetism. In order to avoid trivial boundary effects, we exclude $L/4$ sites from each end of the chain, so that the computed three-point function accurately reflects the bulk behavior. We find that for moderately large $U \,(\ge 5\,\tilde{t})$, $G (r)$ is always negative for $r$\,$=$\,$1$, independent of system size, and never becomes appreciably positive for distances of up to $r=8$. This reveals that the spins tend to be partially antialigned near an excess electron, and the magnitude of this anticorrelation increases with doping. Hence, a ferromagnetic polaron never forms.

\begin{figure*}[tb]
    \centering
    \includegraphics[width=0.95\linewidth]{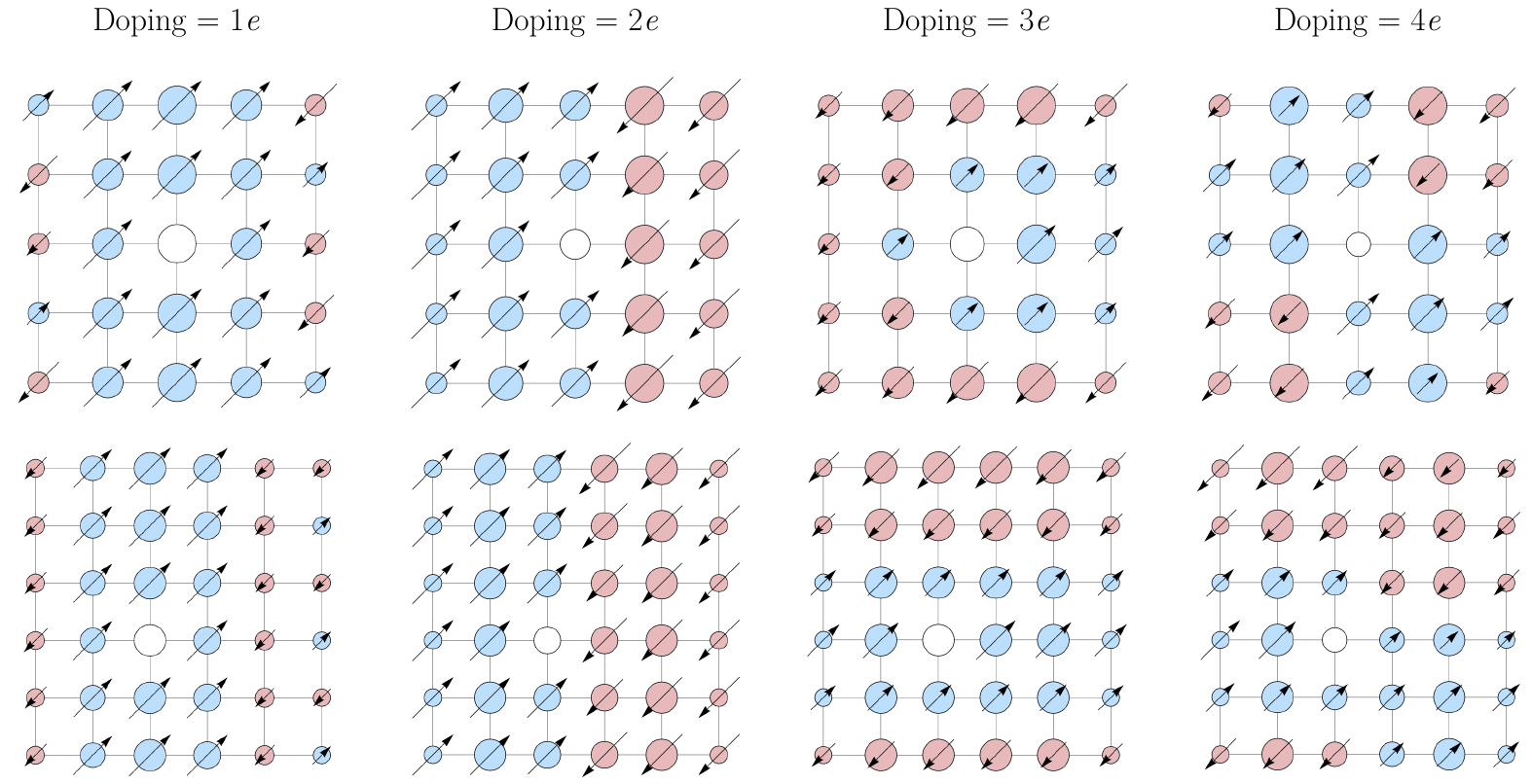}
    \caption{DMRG ground states on $5\times5$ (top) and $6\times6$ (bottom)  square clusters with cylindrical boundary conditions, doped with one to four electrons, for $\tilde{t}/t=4$, $U = 10\,\tilde{t}$. All the plots follow the same conventions as used in Fig.~\ref{fig:OBC}.}
    \label{fig:CBC}
\end{figure*}

Since the site-averaged correlation function $G (r)$ could potentially suffer from cancellations between contributions from electron-rich and hole-rich spatial regions, due to the factor of $(n_i-1)$ in Eq.~\eqref{eq:C3}, we also compute the modified three-point function
\begin{equation}
\label{eq:C3Sq}
    \tilde{G}^{}_i (r) = \left \langle \left(n^{}_i -1\right)^2 \mathbf{S}^{}_{i-r}\cdot\mathbf{S}^{}_{i+r} \right \rangle.
\end{equation}
While not explicitly shown in Fig.~\ref{fig:1Dpolaron}, the site-averaged $G^{}_i (r)$ and $\tilde{G}^{}_i (r)$ are found to be virtually identical, indicating that for the doped system, the dominant contribution to $G(r)$ is from the majority carriers.

\section{Square arrays on cylinders}
\label{sec:CBC}

The ground states of the extended Hubbard model on square clusters with cylindrical boundary conditions are arrayed in Fig.~\ref{fig:CBC}. While these states exhibit some similarities to the ones with open boundaries, displayed in Fig.~\ref{fig:OBC}, a new feature, for certain dopings, is the development of stripe ordering \cite{samajdar2023nagaoka}. Such stripes are well exemplified, for instance, by the two-electron-doped systems, which convey that it can sometimes be energetically favorable to form two smaller ferromagnetic domains (thus optimizing the antiferromagnetic exchange contribution along the long domain wall) at the expense of a single larger one. This is because the periodic boundaries along the circumference of the cylinder increase the kinetic energy gain from delocalization over a given area, relative to a system with open boundaries, and together with the superexchange, this can offset the energetic cost of confining the doublon to a smaller spatial region. Under certain circumstances, such as for the $6\times 6$ system doped with four electrons, the spin texture can also form square domains, as opposed to elongated stripes. However, rather than a simple checkerboard arrangement, we observe that each $3\times 3$ domain is slightly displaced from the one adjacent to it so as to allow the doublons to delocalize over a larger ferromagnetically ordered region.

\end{appendix}

\bibliographystyle{apsrev4-2}
\bibliography{refs.bib}

\end{document}